\documentclass[prb,twocolumn,groupedaddress]{revtex4}

\usepackage{graphicx}
\usepackage{dcolumn}
\usepackage{bm}

\begin{document}

\preprint{ATB-1}

\title{Spin Dynamics of Half-Doped La$_{3/2}$Sr$_{1/2}$NiO$_4$}

\author{P. G. Freeman}
\homepage{http://xray.physics.ox.ac.uk/Boothroyd}
\author{A. T. Boothroyd}
\author{D. Prabhakaran}
\affiliation{Department of Physics, Oxford University, Oxford, OX1 3PU, United
Kingdom }

\author{C. Frost}
\affiliation{ISIS Facility, Rutherford Appleton Laboratory,
Chilton, Didcot, OX11 0QX, United Kingdom}
\author{M. Enderle}
\author{A. Hiess}\affiliation{
Institut Laue-Langevin, BP 156, 38042 Grenoble Cedex 9, France }

\date{\today}

\begin{abstract}

We report polarized- and unpolarized- neutron inelastic scattering
measurements of the magnetic excitation spectrum in the
spin-charge ordered phase of La$_{3/2}$Sr$_{1/2}$NiO$_4$. Up to
energies of $\sim$30\,meV we observe broad magnetic modes
characteristic of a near checkerboard ordering. A linear spin-wave
model for an ideal checkerboard ordering with a single
antiferromagnetic exchange interaction $J' = 5.8 \pm 0.5$\,meV
between next-nearest-neighbour spins on Ni$^{2+}$ sites, together
with a small {\it XY}-like single-ion anisotropy, provides a
reasonable description of the measured dispersion. Above 30\,meV
the excitations are not fully consistent with the linear spin-wave
model, with modes near the two-dimensional reciprocal space
wavevector $(0.5,0.5)$ having an anomalously large intensity.
Furthermore, two additional dispersive modes not predicted by spin
wave theory were observed, both of which are probably magnetic.
One disperses away from $(0.5,0.5)$ in the energy range between
50--56\,meV, and the other appears around ($h$,\ $k$) type
positions ($h,k =$ integer) in the energy range 31--39\,meV. We
propose a model in which these anomalous features are explained by
the existence of discommensurations in the checkerboard ordering.
At low energies there is additional diffuse scattering centred on
the magnetic ordering wavevector. We associate this diffuse
scattering with dynamic antiferromagnetic correlations between
spins attached to the doped holes.

\end{abstract}

\maketitle

\section{\label{sec:intro}Introduction}

The existence of charge ordering phenomena continues to attract
considerable interest, especially concerning the properties of
layered transition-metal oxides. The nature of the charge ordering
in these materials depends on the degree of doping relative to a
Mott insulating phase. At low doping the charges tend to segregate
into parallel lines. The resulting stripe superstructures have
been widely discussed in connection with high-temperature
superconductivity in cuprates.\cite{Tranquada-Nature-1995} At
higher doping there is a competition between charge-ordered and
metallic states. The existence of magnetic correlations is also an
important factor, and the coupling between charge, orbital, spin
and lattice degrees of freedom can often stabilize novel ordered
states.

At half doping, systems with dominant Coulomb repulsions are
expected to exhibit a stable Wigner crystal state, which in the
case of a two-dimensional square lattice takes the particularly
simple form of a checkerboard pattern. This type of charge order
has been observed in the isostructural layered perovskites
La$_{3/2}$Sr$_{1/2}$NiO$_4$,\cite{chen-PRL-1993}
La$_{3/2}$Sr$_{1/2}$MnO$_4$,\cite{LSMO_half} and
La$_{3/2}$Sr$_{1/2}$CoO$_4$,\cite{Zaliznyak-PRL-2000}.
Interestingly, checkerboard charge order has also recently been
observed at a doping levels well below 0.5 in
Bi$_{2}$Sr$_{2}$CaCu$_2$O$_{8+\delta}$,\cite{hoffman-Science-2002},
and Ca$_{2-x}$Na$_{x}$CuO$_2$Cl$_{2}$,\cite{Hanaguri-Nature-2004}
with a periodicity of 4 Cu atoms. Further experimental studies of
simple systems with checkerboard charge order could help explain
the formation of electronically-ordered states in more complex
systems.

In this paper we describe a neutron inelastic scattering study of
the spin excitations in the half-doped layered nickelate
La$_{3/2}$Sr$_{1/2}$NiO$_4$. This compound exhibits checkerboard
charge order below $T_{\rm CO}\approx
480$\,K.\cite{chen-PRL-1993,kajimoto} Below $T_{\rm IC}\approx
180$\,K the checkerboard pattern becomes slightly incommensurate,
and below $T_{\rm SO}\approx 80$\,K it is accompanied by
incommensurate magnetic order.\cite{kajimoto} A spin-reorientation
transition has been observed at $T_{\rm SR}=57$\,K.\cite{me} By
mapping out the spin excitation spectrum of
La$_{3/2}$Sr$_{1/2}$NiO$_4$ as a function of energy and wavevector
and comparing the dynamic response with that of other
charge-ordered compounds we hope to gain a better understanding of
the interactions that stabilize the ground state in these systems.

Previously, the spin excitation spectrum of
La$_{2-x}$Sr$_{x}$NiO$_4$ has been investigated over a wide range
of wavevector and energy in stripe-ordered compounds with $x =
0.275$,\cite{boothroyd-JMMM-2004} $x =
0.31$,\cite{Bourges-PRL-2003} and $x=
1/3$.\cite{boothroyd-PRB-2003,boothroyd-PRL-2003}  The spin
excitations are highly two-dimensional (2D), and fairly
independent of $x$ over this small range of compositions studied.
Over most of the energy range the magnetic dynamics can be
described in terms of spin wave excitations of the
antiferromagnetically ordered Ni spins in the region between the
charge stripes.\cite{boothroyd-PRB-2003} At low energies, however,
a second magnetic component is observed that has been shown to
come from quasi-1D antiferromagnetic spin fluctuations along the
stripes themselves.\cite{boothroyd-PRL-2003}

\begin{figure}[!ht]
\begin{center}
\includegraphics{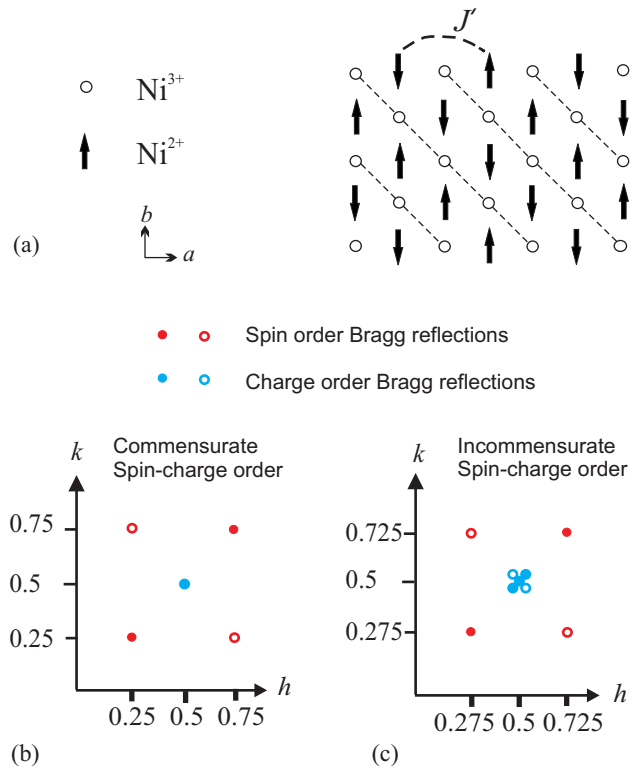}
\caption[Checkerboard and peak positions]{(Color online)(a) Ideal
checkerboard spin--charge ordering in a NiO$_2$ square lattice.
Circles represent Ni$^{3+}$ holes, and solid arrows represent
spins on Ni$^{2+}$ sites. The broken lines are included to
highlight that the spin pattern breaks the 2D symmetry of the
checkerboard charge ordering. $J'$ is the exchange coupling
parameter of the Ni$^{2+}$ spins across the Ni$^{3+}$ site. This
commensurate ordering is not realized in practise in
La$_{3/2}$Sr$_{1/2}$NiO$_{4}$. (b) Diagram of part of the $(h,k)$
plane in 2D reciprocal space showing the positions of the
first-order magnetic and charge order Bragg peaks for the ideal
checkerboard ordering represented in (a). The peaks from the
equivalent domain in which the magnetic ordering is rotated by
90\, deg relative to that in (a) are superimposed. (c) The same
diagram as (b) except with the magnetic and charge order Bragg
peaks observed in the incommensurate ordered phase of
La$_{3/2}$Sr$_{1/2}$NiO$_{4}$. For simplicity we neglect the
variation in the peak positions in the direction perpendicular to
the NiO$_2$ plane. } \label{fig:fig1}
\end{center}
\end{figure}

Figure\ \ref{fig:fig1} illustrates how the different ordered
phases in La$_{3/2}$Sr$_{1/2}$NiO$_4$ can be identified in
diffraction measurements. Figure\ \ref{fig:fig1}(a) is a
simplified model of the ground state spin--charge order within the
NiO$_2$ layers of La$_{3/2}$Sr$_{1/2}$NiO$_4$ neglecting the small
incommensurate modulation. For present purposes we assume the
doped holes are site-centred, so that the checkerboard pattern is
derived from alternating Ni$^{2+}$ and Ni$^{3+}$ ions carrying
spins $S=1$ and $S=\frac{1}{2}$, respectively. The $S=1$ spins are
assumed to be ordered antiferromagnetically, but we make no
assumption at this stage about magnetic order of the
$S=\frac{1}{2}$ spins (this issue will be addressed later in the
paper). The positions in reciprocal space of the corresponding
Bragg peaks are shown in Fig.\ \ref{fig:fig1}(b). Peaks from the
charge order have two-dimensional wave vectors (in units of
$2\pi/a$) $(h+\frac{1}{2},k+\frac{1}{2})$, where $h$ and $k$ are
integers.\cite{notation} The magnetic order has double the
periodicity of the charge order, so peaks from the magnetic order
appear at $(h+\frac{1}{2},k+\frac{1}{2}) \pm (\frac{1}{4},
\frac{1}{4})$. Rotation of the ordering pattern by 90\,deg
generates an equivalent magnetic structure, this time with
magnetic peaks at $(h+\frac{1}{2},k+\frac{1}{2}) \pm (\frac{1}{4},
-\frac{1}{4})$. In the absence of a symmetry-breaking interaction
we expect an equal population of these two domains, so the pattern
of Bragg peaks will be a superposition, as shown in Fig.\
\ref{fig:fig1}(b).

The actual spin--charge ordered phase of
La$_{3/2}$Sr$_{1/2}$NiO$_4$ observed below $T_{\rm IC}$ does not
conform to the ideal structure shown in Fig.\ \ref{fig:fig1}(a).
Instead, the magnetic Bragg peaks are found at the incommensurate
positions $(h+\frac{1}{2},k+\frac{1}{2}, l) \pm (\epsilon/2,
\epsilon/2, 0)$ with $l$ an odd integer, and
$(h+\frac{1}{2},k+\frac{1}{2}, l) \pm (\epsilon/2, -\epsilon/2,
0)$ with $l$ an even integer, where $\epsilon \approx 0.44$.
\cite{kajimoto,yoshizawa-PRB-2000,me} New charge-order satellite
peaks appear at $(h\pm \epsilon,k\pm \epsilon)$ in addition to the
checkerboard charge order peak at $(h+\frac{1}{2},k+\frac{1}{2})$,
with little or no $l$ dependence.\cite{kajimoto,me} The full set
of 2D magnetic and charge order wavevectors for the incommensurate
phase of La$_{3/2}$Sr$_{1/2}$NiO$_4$, including those for the 90\,
deg domain, is shown in Fig.\ \ref{fig:fig1}(c).

Kajimoto {\it et al.}\cite{kajimoto} suggested that two types of
charge order coexist in the incommensurate phase, one part of the
system being charge-ordered in a checkerboard pattern and the
other part adopting an incommensurate, stripe-like, spin-charge
order. These authors developed models for the latter component
based on the introduction of diagonal discommensurations in the
ideal checkerboard spin-charge structure of Fig.\
\ref{fig:fig1}(a). As pointed out by Kajimoto {\it et al.}, the
stability of the incommensurate structure is probably the result
of a competition between magnetic and electrostatic energy. The
strong superexchange interaction favours having antiparallel spins
on nearest neighbour Ni sites, whereas the Coulomb interaction
tries to have a uniform charge density.

\section{\label{sec:exper}Experimental details}

The neutron scattering measurements were performed on a single
crystal of La$_{3/2}$Sr$_{1/2}$NiO$_{4}$ grown by the
floating-zone method.\cite{Prab} The crystal was a cylinder 35\,mm
in length and 7\,mm in diameter.  The oxygen content of the
crystal was determined by thermogravimetric analysis (TGA) to be
$4.02\pm 0.01$. This is the same crystal that was used in our
earlier neutron diffraction and magnetization study of the
magnetic order.\cite{me}

The majority of the unpolarized-neutron scattering measurements
were performed on the time-of-flight chopper spectrometer MAPS at
the ISIS Facility. The crystal was mounted on MAPS in a
closed-cycle refrigerator and aligned with the $c$ axis parallel
to the incident beam direction. A Fermi chopper was used to select
the incident neutron energy. Incident energies of 60 and 100\,meV
were used. The intensity was normalized and converted to units of
scattering cross-section
(mb\,sr$^{-1}$\,meV$^{-1}$\,[f.u.]$^{-1}$) by comparison with
measurements from a standard vanadium sample. Scattered neutrons
were recorded in large banks of position-sensitive detectors. The
spin dispersion was found to be highly two dimensional. Hence, we
analyzed the data by making a series of constant energy slices and
projecting the intensities onto the $(h, k)$ two-dimensional
reciprocal lattice plane.

Further unpolarized- and polarized-neutron measurements were
performed on the triple-axis spectrometers IN8 and IN20 at the
Institut Laue-Langevin. The energies of the incident and scattered
neutrons were selected by Bragg reflection from a double-focusing
bent Si crystal monochromator on IN8, and an array of Heusler
alloy crystals on IN20. On both instruments the data were obtained
with a final neutron wavevector of 2.66\,\AA$^{-1}$, and a
pyrolytic graphite (PG) filter was placed between the sample and
analyzer to suppress higher-order harmonic scattering. For
polarized-neutron scattering on IN20 the spin polarization, ${\bf
P}$, was maintained in a specified orientation with respect to the
neutron wavevector, ${\bf Q}$, by an adjustable guide field of a
few mT at the sample position. For the experiment on IN8 the
crystal was orientated with the [100] and [010] crystal directions
in the horizontal scattering plane, so that ($h$,\ $k$,\ 0)
positions in reciprocal space could be accessed. On IN20, we
mounted the crystal with the [001] and [110] directions in the
horizontal scattering plane, so that ($h$,\ $h$,\ $l$) positions
in reciprocal space could be accessed.

\section{\label{sec:exper}Results}

\begin{figure}[!ht]
\begin{center}
\includegraphics[width=8cm,clip=]{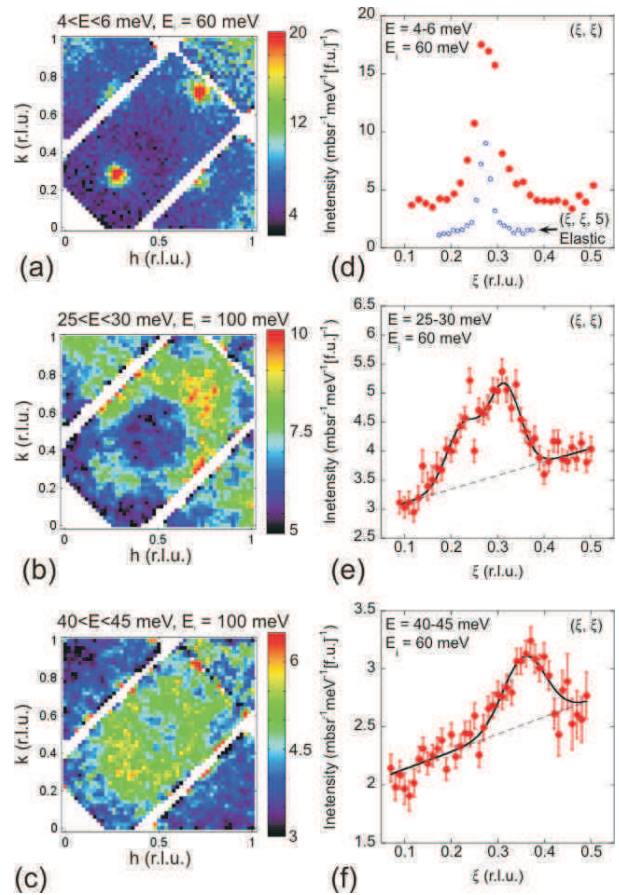}
\caption[cuts]{(Color online) Neutron scattering measurements from
La$_{3/2}$Sr$_{1/2}$NiO$_{4}$ measured on MAPS at $T =10$\,K.
$(a)-(c)$ are constant-energy slices showing the variation of the
intensity in the $(h,k)$ plane at different energies. The data are
averaged over the range of energies indicated above the figures.
The incident neutron energy was 60\,meV for (a), and 100\,meV for
(b) and (c). Data from four equivalent Brillouin zones have been
averaged. $(d)-(f)$ show cuts along the $(\xi,\xi)$ direction for
the same energies as shown in $(a)-(c)$. The solid lines are the
results of fits with one Gaussian [(d) and (f)] or two Gaussians
[(e)] on a linear background. In $(d)$ we also display the
$(0.275, 0.275, 5)$ magnetic order Bragg peak measured on IN20.
The out of plane wavevector for the position $(0.275, 0.275)$ are
as follows $(a) \& (d)$ is $l = 0.55$, $(b)\ l = 2.3$, $(e)\ l =
3.0$, $(c)\ l = 3.45$ and $(f)\ l = 5.2$.} \label{fig:cuts}
\end{center}
\end{figure}

Figures \ref{fig:cuts}(a)$-$(c) show constant-energy slices from
runs performed at $T = 10$\,K on MAPS.\cite{ldep} The slices have
been averaged over a range of energies, as indicated in the
figures. In (a), the range is 4--6\,meV, and the intensity is seen
to be enhanced at the four equivalent magnetic order wavevectors
$(0.5, 0.5) \pm (\epsilon/2,\epsilon/2)$ and $(0.5, 0.5) \pm
(\epsilon/2,-\epsilon/2)$, where $\epsilon \approx 0.44$. This
signal, therefore, corresponds to low energy spin excitations from
the magnetically ordered ground state. As the energy increases,
the intensity spreads out but initially remains centred on the
magnetic wavevectors. This can be seen in Fig.\ \ref{fig:cuts}(b),
which shows data averaged over 25--30\,meV. At still higher
energies the four blobs of intensity tend to merge towards $(0.5,
0.5)$. This is illustrated in (c), which corresponds to an energy
range 40--45\,meV.

In Figs.\ \ref{fig:cuts}(d)--(f) we plot cuts through the data in
the $(\xi,\xi)$ direction for the same energy ranges as used in
Figs.\ \ref{fig:cuts}(a)--(c).  At low energies, $E = 4 - 6$\,
meV, the scattering takes the form of a single peak centred on the
magnetic wavevector. The fitted Gaussian half-width of this peak
converts to a correlation length of $35 \pm 1$\,\AA (the inverse
of the half width at half maximum). At energies of 25--30\,meV
[Fig.\ \ref{fig:cuts}(e)] the lineshape is also centred on the
magnetic wavevector, but it now contains two peaks which can just
be resolved [another example can be seen in Fig.\
\ref{fig:Ering}(g)]. The fitted width of these peaks corresponds
to a correlation length of $24 \pm 2$\,\AA. As the energy
increases, the right-hand of the two peaks grows while the
left-hand peaks diminishes. Above $E \approx 40$\,meV the
left-hand peak has virtually no intensity, as can be seen in Fig.\
\ref{fig:cuts}(f).

\begin{figure}[!ht]
\begin{center}
\includegraphics[width=8cm,clip=]{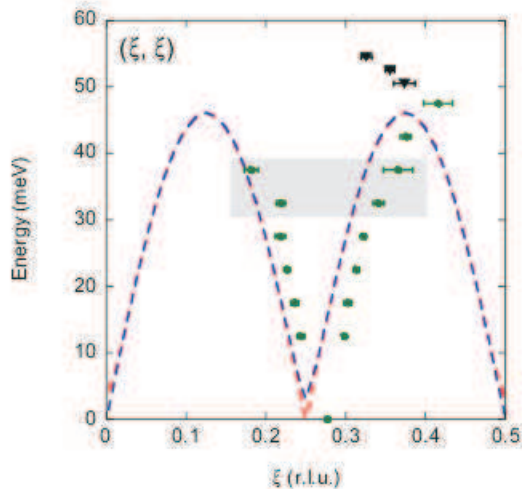}
\caption[Dispersion]{(Color online) Dispersion of the magnetic
excitations in La$_{3/2}$Sr$_{1/2}$NiO$_{4}$ parallel to the
$(\xi,\ \xi)$ direction. The points are the results of fits to
cuts such as those shown in Fig.\ \ref{fig:cuts}. The broken line
is calculated from Eq.\ (\ref{eq:dispersion}), the spin wave
dispersion for a checkerboard ordered system with exchange
parameter $J' = 5.8$\,meV and out-of-plane anisotropy $K_ c =
0.05$\,meV. Triangles above the calculated dispersion curve
indicate the positions of an additional observed scattering mode.
The shaded area represents the region in which additional
scattering hampers the study of the spin-wave excitations, see
Fig.\ \ref{fig:Ering}.} \label{fig:dispersion}
\end{center}
\end{figure}

Figure \ref{fig:dispersion} shows the dispersion of the spin
excitations obtained from Gaussian fits to the peaks observed in
cuts such as those displayed in Fig.\ \ref{fig:cuts}(d)--(f). The
points in Fig.\ \ref{fig:dispersion} are the fitted peak centres
corrected for the effect of the non-zero width of the cut
perpendicular to the cut direction.  For this purpose the
scattering was taken to be a circle centred on the magnetic
wavevector. Within the experimental limits of our data the
dispersion was found to be the same in orthogonal cut directions.

\begin{figure}[!t]
\begin{center}
\includegraphics[width=8cm,clip=]{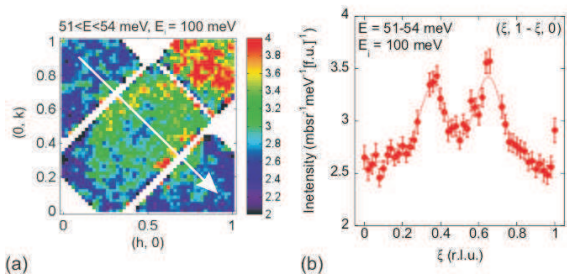}
\caption[slice]{(Color online) Neutron scattering measurements
from La$_{3/2}$Sr$_{1/2}$NiO$_{4}$ measured on MAPS at $T =10$\,K.
(a) Constant-energy slice through the data showing the intensity
distribution in the $(h,k)$ plane in the energy range 51--54\,meV.
The slice reveals weak diffuse scattering symmetrically
distributed around $(0.5, 0.5)$. Data from four equivalent
Brillouin zones have been averaged to improve the statistics. (b)
Cut through the 51--54\,meV slice along the path indicated by the
arrow in (a).} \label{fig:ring}
\end{center}
\end{figure}

In constructing the dispersion curve we had to take care to avoid
confusing the scattering from magnetic excitations and phonons,
especially in the energy range 10--40\,meV where the phonon
scattering is particularly strong. One check we made was to
compare data obtained at 300\,K with that obtained at 10\,K.
Phonon scattering increases in strength with temperature, whereas
magnetic scattering decreases in strength above the magnetic
ordering temperature.

At energies above 50\,meV we observed a broad ring of scattering
centred on $(0.5, 0.5)$. This is illustrated in Fig.\
\ref{fig:ring}, which shows (a) a constant-energy slice averaged
over the energy range 51--54\,meV, and (b) a cut through this data
in the $(\xi,1-\xi)$ direction. This mode disperses away from
$(0.5, 0.5)$ with increasing energy, but was too weak to measure
above 56\,meV. This scattering was found to be slightly weaker at
300\,K than at 10\,K, $34\pm19$\,
origin for it. We have added the peak positions of this scattering
to the dispersion curve in Fig.\ \ref{fig:dispersion}.

\begin{figure}[!t]
\begin{center}
\includegraphics[width=8cm,clip=]{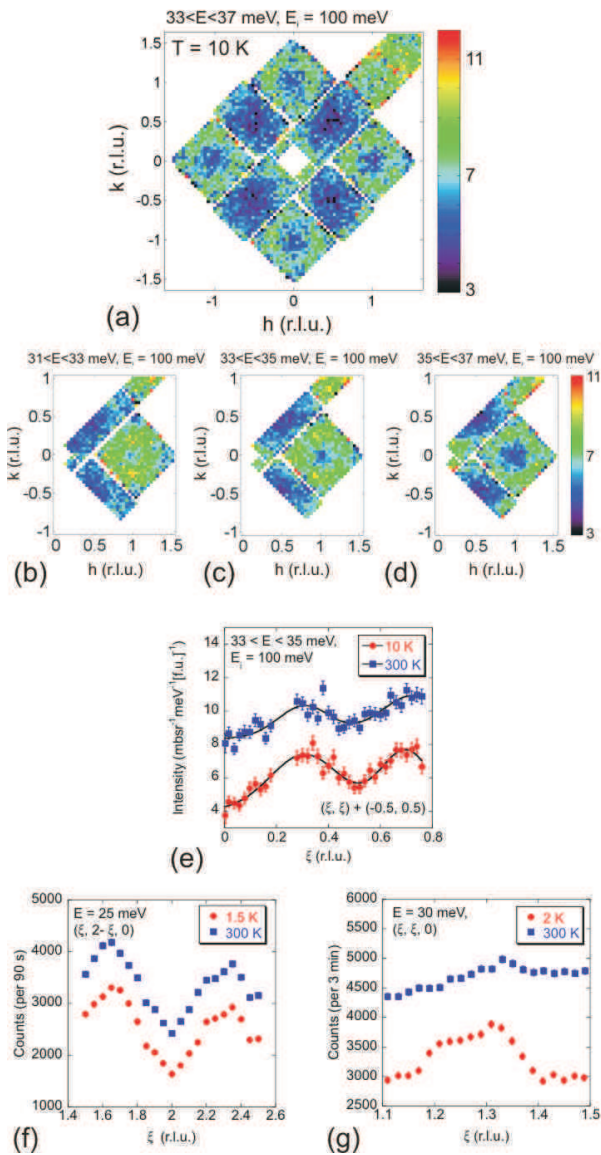}
\caption[slice]{(Color online) Excitations from
La$_{3/2}$Sr$_{1/2}$NiO$_{4}$. (a) Distribution of scattering
intensity in the $(h,\ k)$ plane measured on MAPS at $T = 10$\,K.
The scattering has been averaged over the energy range $E = 33 -
37$\,meV. (b)--(d) Slices through the scattering intensity
measured at $T = 10$\,K in the $(h\ ,k)$ plane averaged over the
energy ranges (b) $31 - 33$\,meV, (c) $33 - 35$\,meV, (d) $35 -
37$\,meV. The data in the MAPS detector have been folded to take
advantage of the symmetry of the scattering. (e) Cuts through the
data shown in (c) along a line parallel to the $(h,\ h)$ direction
passing through $(1,\ 0)$. The data at $T = 10$\,K and $300$\,K,
have been offset by 5 units. The lines in (e) are the result of
fitting the points with two gaussian functions on a sloping
background. (f) Scans of the scattering observed at $E = 35$\,meV
for $T = 10$\,K and $300$\,K taken on IN8. (g) A scan through the
spin wave dispersion at $E= 30$\,meV for $T = 10$\,K and $300$\,K
taken on IN8. The 300\,K data has been offset by the addition of
1000 counts. } \label{fig:Ering}
\end{center}
\end{figure}

In the energy range $31-39$\,meV we observed an interesting
scattering feature separate from the main spin wave scattering.
Figure \ref{fig:Ering}(a) shows the distribution of intensity
measured on MAPS for energies averaged over the range 33--37\,meV
covering a large area of reciprocal space. This intensity map
displays strong diffuse scattering rings centred on the reciprocal
lattice zone centres $(h,k)$, where $h$ and $k$ are integers.
Figures \ref{fig:Ering}(b)--(d) present constant-energy slices
through this scattering centred on energies of 32\,meV, 34\,meV
and 36\,meV. In these slices the symmetry-equivalent data have
been folded into one quadrant to improve the statistics. The
intensity is seen to disperse away from the $(1, 0)$ point. In
these measurements the out-of-plane component of the scattering
vector varies from $l=$ 2.5 to 4. A search was made with the IN8
spectrometer of other Brillouin zones in the $(h, k, 0)$
reciprocal lattice plane. This survey confirmed the results found
on MAPS, showing that the scattering was not restricted to a
particular out-of-plane component of the scattering vector.

To assess whether or not this scattering is magnetic we performed
measurements as a function of temperature. Cuts through the MAPS
data at 10\,K and 300\,K are shown in Fig. \ref{fig:Ering}(e).
These data show the scattering to decrease slightly with
temperature, being approximately $45\pm12\%$ stronger at 10\,K
than at 300\,K. We also observed similar scattering in the range
25--30\,meV. Fig.\ref{fig:Ering}(f) shows ${\bf Q}$ scans through
the point $(2,0,0)$ at an energy of 25\,meV for temperatures of
1.5\,K and 300\,K, measured on IN8. The scattering here appears to
be temperature independent. To understand these temperature
effects we need to take into account that this diffuse scattering
ring overlaps the magnetic ordering wavevectors, so will contain
some spin-wave scattering. The spin wave scattering itself will,
of course, be temperature dependent. To illustrate this we show in
Fig.\ \ref{fig:Ering}(g) scans through the spin-wave scattering
associated with the magnetic wavevector $(1.5,1.5)
-(\epsilon/2,\epsilon/2)$ measured at a slightly higher energy
($E=30$\,meV) on IN8 at $T = 10$\,K and 300\,K. The amplitude of
the spin wave scattering clearly decreases with temperature. From
this loss of intensity we can infer the temperature dependence of
the remaining component of the scattering in Figs.\
\ref{fig:Ering}(a)--(f). Our analysis indicates that at $E \sim
35$\,meV the non-spin-wave component of the diffuse scattering
ring presented in Figs.\ \ref{fig:Ering}(a)--(d) decreases in
intensity with temperature, whereas at $E= 25$\,meV [Fig.\
\ref{fig:Ering}(f)] it increases with temperature. This leads us
to conclude that the diffuse scattering around the Brillouin zone
centres shown in Figs.\ \ref{fig:Ering}(a)--(d) is most likely
magnetic in origin.

Further support for this conclusion was provided by examining data
we have collected on crystals of La$_{2-x}$Sr$_{x}$NiO$_4$ with
$x=1/3$ and $0.275$ under similar condition during a separate
experiment on MAPS.\cite{Woo} Constant-energy slices for $x=1/3$
and $0.275$ in the energy range $31-39$\,meV do not show any
diffuse scattering rings like those shown in Figs.\
\ref{fig:Ering}(a)--(d), so this feature seems to be specific to
$x=1/2$. This makes it highly unlikely that this scattering comes
from phonons intrinsic to the host lattice of
La$_{2-x}$Sr$_{x}$NiO$_4$.

For energies below 30\,meV we performed additional measurements
with polarized neutrons. In this energy range there is strong
scattering from phonons, and polarized neutrons were necessary to
provide an unambiguous separation of the magnetic and non-magnetic
scattering. We were particularly interested in studying the
magnetic scattering as a function of energy because earlier
measurements on La$_{2-x}$Sr$_{x}$NiO$_4$ compositions with
$x=0.275$, 0.33 and 0.37 had revealed unexpected structure in the
energy spectrum.\cite{boothroyd-PRB-2003,boothroyd-PhysicaB} For
these measurements the neutron polarization ${\bf P}$ was aligned
parallel to the scattering vector ${\bf Q}$ so that the spin-flip
(SF) channel would contain purely magnetic scattering.

\begin{figure}[!ht]
\begin{center}
\includegraphics{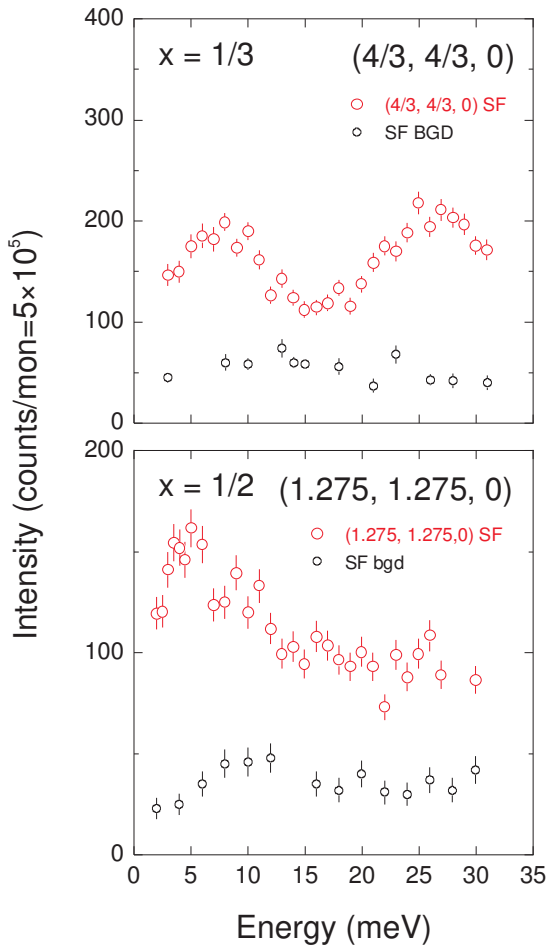}
\caption[Escans]{(Color online) Polarized-neutron scattering from
charge-ordered La$_{2-x}$Sr$_{x}$NiO$_{4}$ with $x = 1/3$ and 1/2.
The plots show energy scans performed at the magnetic ordering
wavevector appropriate to each composition. The spin-flip (SF)
scattering was measured with {\bf P} $\parallel$ {\bf Q}, and the
SF background was estimated from scans centred away from the
magnetic ordering wavevector. The measurements were made at 10\,K
for $x = 1/2$ and 13.5\,K for $x = 1/3$.} \label{fig:polar}
\end{center}
\end{figure}

The most interesting finding is reproduced in Fig.\
\ref{fig:polar}, which shows energy scans measured in the SF
channel on samples with compositions $x=1/3$ and $x=1/2$ at their
respective magnetic ordering wavevectors. The $x=1/3$ data
contains two peaks, one centred on 7\,meV and the other on
26\,meV. The $x=1/2$ data contains a single peak at 5\,meV. The
lower energy peaks were shown to be gaps due to spin
anisotropy.\cite{boothroyd-PRB-2003,boothroyd-PhysicaB} Below
these gaps the out-of-plane spin fluctuations are quenched. The
origin of the higher energy peak in the $x=1/3$ data, which was
first discussed in Ref. \onlinecite{boothroyd-PRB-2003} and which
was also found in similar data from crystals with $x=0.275$ and
$x=0.37$,\cite{boothroyd-PhysicaB} remains a mystery. Of all the
compositions studied, the one with $x=1/3$ exhibits this feature
most strongly. The absence of a corresponding peak in the $x=1/2$
data seems to suggest that this peak is a property of compounds
with static stripe order with a periodicity of $\sim$3 lattice
spacings. The monotonic decrease in intensity above 5\,meV in the
$x=1/2$ data is qualitatively consistent with the expected $1/E$
dependence of the cross section for scattering from
antiferromagnetic spin waves.

\begin{figure}[!t]
\begin{center}
\includegraphics[width=8cm,clip=]{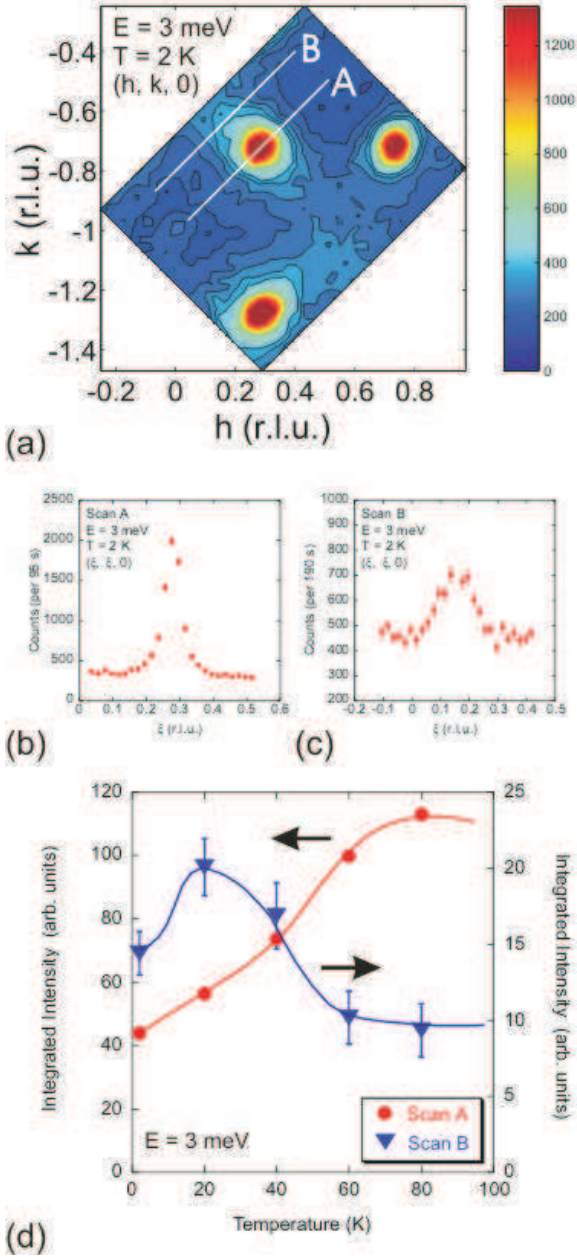}
\caption[slice]{(Color online) Low energy scattering from
La$_{3/2}$Sr$_{1/2}$NiO$_{4}$ measured on IN8 at an energy of
3\,meV and a temperature of 2\,K. (a) Map of the scattering
intensity in the $(h, k, 0)$ reciprocal lattice plane. (b) and (c)
Cuts along the lines marked A and B in (a). (d) Temperature
dependence of the integrated intensity of the peaks shown in (b)
and (c). The lines are guides to the eye.} \label{fig:ridge}
\end{center}
\end{figure}

Finally, we describe a feature observed in the scattering from
La$_{3/2}$Sr$_{1/2}$NiO$_{4}$ at low energies.  Figure
\ref{fig:ridge}(a) is a map of the intensity measured on IN8
covering part of the $(h,k,0)$ plane in reciprocal space. The map
was constructed from a series of scans performed parallel to $(h,
h, 0)$ at an energy of 3\,meV and a temperature of 2\,K. Strong
scattering can be seen centred on the magnetic ordering
wavevectors, but additional weak diffuse scattering can also be
seen centred on these same positions. This diffuse scattering is
slightly elongated in the diagonal directions, parallel to the
discommensuration lines in the distorted checkerboard structure.
There was no observable elastic diffuse scattering, so this
feature represents a short-range dynamic magnetic correlation. We
were able to follow the diffuse inelastic scattering up in energy
to $\sim$10\,meV.

In Fig.\ \ref{fig:ridge}(b) and (c) are shown scans through the
spin wave scattering and the diffuse scattering made along the
lines marked A and B, respectively, in Fig.\ \ref{fig:ridge}(a).
Figure \ref{fig:ridge}(d) displays the temperature dependence of
the integrated intensity of the peaks in these two scans. The
intensity for scan A is seen to increase with increasing
temperature. This is due partly to thermal population of the spin
waves, and partly to the reorientation of the ordered moment of
the Ni$^{2+}$ spins that takes place at 57\,K.\cite{me} Scan B, on
the other hand, shows an initial increase in intensity on warming
which peaks at around 20\,K before decreasing at higher
temperatures. This decrease suggests that the diffuse scattering
is magnetic in origin, and the striking difference in temperature
dependence between the spin wave scattering and the diffuse
scattering strongly indicates that they arise from two different
magnetic components in the system.

\section{\label{sec:conc}Discussion}

Let us first summarize the key observations, and then provide some
quantitative analysis. From the various neutron scattering
measurements we have performed on spin- and charge-ordered
La$_{3/2}$Sr$_{1/2}$NiO$_{4}$ we have been able to identify four
distinct features: (1) dispersive spin excitation associated with
the magnetic ordering wavevector; (2) low energy diffuse spin
dynamics also associated with the magnetic ordering wavevector but
with a distinct temperature dependence; (3) a probable magnetic
mode dispersing from $(0.5, 0.5)$ at energies in the range
50--56\,meV; (4) a probable magnetic mode dispersing from $(h, k)$
type positions in the energy range 31--39\,meV. All these features
are relatively broad in wavevector, and therefore arise from
dynamic correlations that are short-range in nature.

An obvious starting point for any analysis of the magnetic
excitations in La$_{3/2}$Sr$_{1/2}$NiO$_{4}$ is the ideal
checkerboard spin--charge ordering pattern shown in Fig.\
\ref{fig:fig1}(a). As far as the spins are concerned, this is a
simple two-sublattice antiferromagnet built from $S=1$ spins on
the Ni$^{2+}$ sites. Spins attached to the Ni$^{3+}$ sites are
ignored for the time being. Following our previous work on
stripe-ordered
La$_{5/3}$Sr$_{1/3}$NiO$_{4}$,\cite{boothroyd-PRB-2003} we adopt a
spin Hamiltonian of the form
\begin{equation}
H = J' \sum_{\langle i,j\rangle} {\bf S}_i \cdot {\bf S}_j + K_c
\sum_{i} ({\bf S}_i^z)^2,\label{eq:linear}
\end{equation}
where the first summation describes the exchange interactions
between pairs of Ni$^{2+}$ spins in linear
Ni$^{2+}$--O--Ni$^{3+}$--O--Ni$^{2+}$ bonds, and the second
summation describes the small, $XY$-like, single-ion anisotropy.
Here, as in Ref.\ \onlinecite{boothroyd-PRB-2003}, $J'$ is defined
as the exchange energy {\it per spin} (multiply by 2 to obtain the
exchange energy {\it per bond}). We neglect the diagonal exchange
couplings between Ni$^{2+}$ sites which are needed to stabilize
the spin arrangement but are assumed to be small relative to $J'$.
In effect, therefore, the system is treated as 2 uncoupled
square-lattice antiferromagnets with lattice parameter $2a$.

The magnon dispersion derived from Eq.~(\ref{eq:linear}) in the
linear approximation is given by
\begin{equation}
E({\bf Q}) = 8J'S\{(1+K_c/8J')^2-[\gamma({\bf Q})\pm
K_c/8J']^2\}^{1/2}, \label{eq:dispersion}
\end{equation}
where
\begin{equation}
\gamma({\bf Q}) =
\frac{1}{2}[\cos(2Q_xa)+\cos(2Q_ya)].\label{eq:gamma}
\end{equation}
The splitting of the two branches of the dispersion curve is such
that at the magnetic zone centre one mode is gapped and the other
isn't. The size of the gap is $4S(2J'K_c)^{1/2}$. When $K_c \ll
J'$ the maximum energy of the dispersion curve is approximately
$8J'S$.

In Fig.\ \ref{fig:dispersion} we have plotted the spin wave
dispersion along the $(\xi,\xi)$ direction calculated from Eq.\
(\ref{eq:dispersion}) with $S=1$, $J' = 5.8$\,meV and $K_ c =
0.05$\,meV. These parameters were chosen to match the observed
maximum spin wave energy ($\sim 45$\,meV) and anisotropy gap
($\sim 3$\,meV). The spin-wave dispersion curve is seen to provide
a reasonable description of the experimental data, apart from the
obvious shift from the observed incommensurate wavevector $(0.275,
0.275)$ to the ideal checkerboard wavevector of $(0.25,0.25)$.
There is no detectable scattering from spin wave modes on the
dispersion curve near $\xi=0$ and $\xi=0.5$ because the
antiferromagnetic structure factor is small in the magnetic
Brillouin zones centred on $(0,0)$ and $(0.5,0.5)$.

From this analysis we can give a rough estimate of the exchange
and anisotropy parameters for La$_{3/2}$Sr$_{1/2}$NiO$_{4}$. After
consideration of the experimental errors, these are $J' = 5.8 \pm
0.5$\,meV and $K_c = 0.05 \pm 0.02$\,meV. It is interesting to
compare these values with those derived from similar spin-wave
analyses performed on other La$_{2-x}$Sr$_{x}$NiO$_{4}$
compositions. For $x=1/3$ the exchange parameters were found to be
$J = 15\pm 1.5$\,meV, $J' = 7.5 \pm 1.5$\,meV and $K_c = 0.07 \pm
0.01$\,meV,\cite{boothroyd-PRB-2003} where $J$ is the exchange
interaction between Ni$^{2+}$ spins on nearest-neighbour lattice
sites. For undoped La$_2$NiO$_4$ the results were $J = 15.5$\,meV
and $K_c = 0.52$\,meV.\cite{Nakajima-JPSJ-1993} This comparison
shows that $J'$ and $K_c$ are similar for $x=1/3$ and $x=1/2$, but
that $K_c$ is very much larger in undoped La$_2$NiO$_4$. An
explanation for why the single-ion anisotropy reduces so
dramatically with doping is so far lacking.

We now discuss some of the obvious shortcomings of the model. We
have already mentioned that the magnetic ordering wavevector in
La$_{3/2}$Sr$_{1/2}$NiO$_{4}$ is ${\bf q}_{\rm m} = (0.275,0.275)$
not $(0.25,0.25)$. Other problems are: (1) the spin-wave
scattering intensity above $\sim 25$\,meV becomes progressively
more asymmetric about ${\bf q}_{\rm m}$ with increasing energy,
i.e.\ stronger on the side nearest to $(0.5,0.5)$ --- see Figs.\
\ref{fig:cuts}(e) and (f), and Fig.\ \ref{fig:Ering}(g). This
disagrees with the spin wave theory for the model described above,
which predicts a symmetric scattering intensity about the magnetic
zone centre. (2) The extra scattering intensity observed around
$(0.5,0.5)$ above 50\,meV, and around (1,0) and equivalent
positions in the energy range $31-39$\,meV suggest extra magnetic
modes not present for a checkerboard ordering. (3) The spin wave
scattering is very broad, implying that the spin waves propagate
only a few lattice spacings before being scattered or decaying
into another excitation channel. (4) The source of the low energy
diffuse scattering needs to be identified.

To explain these features unequivocally we need a complete
description of the static ordering, which we don't have. However,
the idea that the static incommensurate order in
La$_{3/2}$Sr$_{1/2}$NiO$_{4}$ can be understood in terms of an
ideal checkerboard pattern broken up periodically by
discommensurations is physically appealing,\cite{kajimoto} and
might provide some clues as to the origin of the various features
in the spin excitation spectrum.

\begin{figure}[!ht]
\begin{center}
\includegraphics[width=8cm,clip=]{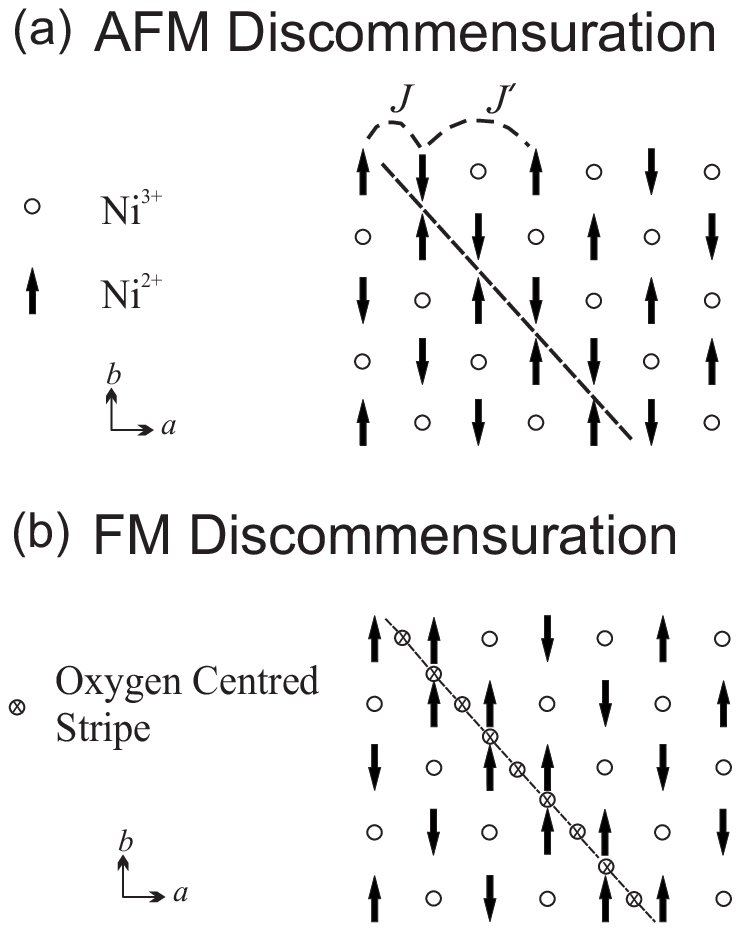}
\caption[Dispersion]{(Color online) Two types of discommensuration
that could exist in La$_{3/2}$Sr$_{1/2}$NiO$_{4}$.\cite{kajimoto}
Circles and arrows denote holes and $S=1$ spins, respectively, on
the Ni sites, in (a) the couplings between next nieghbour
Ni$^{2+}$ spins, $J$, and Ni$^{2+}$ spins across the Ni$^{3+}$
sites, $J'$, are indicated. The circled x symbols show the
positions of an oxygen centred charge stripe. The spins on nearest
neighbour Ni sites have (a) antiparallel and (b) parallel
alignment, respectively, in the discommensuration (indicated by a
dashed line). In (b) holes occupy oxygen sites in the
discommensuration. } \label{fig:discomm}
\end{center}
\end{figure}

The two simplest types of discommensuration in the checkerboard
pattern are illustrated in Fig.\ \ref{fig:discomm}. The first
[Fig.\ \ref{fig:discomm}(a)] has a line of nearest-neighbour
antiparallel spin pairs coupled by a superexchange interaction $J$
expected to be close to 15\,meV as found in the $x=0$ and $x=1/3$
compounds. Around the discommensuration the local hole density is
below average. The second [Fig.\ \ref{fig:discomm}(b)] has a line
of holes on oxygen sites, which increases the local hole density
and forces parallel alignment on neighbouring Ni spins through the
double exchange mecahanism. We will refer to these as
antiferromagnetic (AFM) and ferromagnetic (FM) discommensurations.
More complex discommensurations involving a greater degree of
perturbation of the ideal checkerboard pattern are also possible.

Discommensurations provide a mechanism for producing an
incommensurate spin density wave modulation of the checkerboard
pattern, as discussed earlier. At low energies, the magnon
dispersion is expected to be similar to that of the ideal
checkerboard antiferromagnetic ordering, except that the magnon
dispersion is shifted away from $(0.25,0.25)$ to the
incommensurate wavevector, as observed.

Let us focus, however, on the specific structure of the
discommensurations themselves, which is expected to influence the
spin excitation spectrum at higher energies. The AFM
discommensuration can be regarded as a zig-zag chain with AFM
intra-chain exchange $J$ coupled to the checkerboard AFM
background. The latter has exchange $J'$, and $J\approx 2J'$.
Above the maximum energy of the spin wave dispersion ($\sim
45$\,meV) we might expect to observe spin excitations
characteristic of an AFM zig-zag chain. Because there will be
discommensurations running along both diagonals the scattering
will take the form of a square of scattering dispersing away from
$(0.5, 0.5)$ type positions, where the structure factor of the AFM
zig-zag chain is largest. This could explain the observed `ring'
of scattering apparently dispersing from $(0.5,0.5)$ at energies
above 50\,meV. At lower energies the spin excitations will have
mixed checkerboard and zig-zag chain character, and this could
explain why the spin wave scattering becomes stronger on the side
nearest to $(0.5,0.5)$.

Similarly, FM discommensurations resemble FM zig-zag chains. Spin
excitations with FM zig-zag chain character are expected to be
observable near to the structural zone centres. The strength of
the intra-chain double exchange is not known, but this effect
could account for the mode dispersing from $(h,k)$ type positions
observed in the energy range 31--39\,meV.

Discommensurations could also be responsible for the substantial
widths of the spin wave modes. In a region of commensurate
checkerboard order a spin wave can propagate freely, but when it
encounters a discommensuration the uniformity of the magnetic
order is interrupted sharply, which could scatter the spin wave.
In this case, the correlation length of the spin wave peaks should
correspond roughly to the average spacing between
discommensurations. Earlier we found the correlation length to be
$24 \pm 2$\,\AA\  for energies in the range 25--30\,meV, which
corresponds to approximately $9a/\sqrt{2}$, i.e.\ 9 Ni positions
when projected along the diagonal of the square lattice, which is
the distance between discommensurations according to the model
proposed by Kajimoto {\it et al.}.\cite{kajimoto} Further support
for the broadening mechanism proposed here can be found from a
comparison of the widths of spin wave peaks observed in neutron
scattering measurements of La$_{2-x}$Sr$_{x}$NiO$_{4}$ with
$x=0.275$ and $x=1/3$.\cite{boothroyd-JMMM-2004} The $x=1/3$ spin
wave peaks show no measurable broadening, whereas the $x=0.275$
peaks are broadened. This is consistent with our broadening
mechanism since the stripe order of $x=1/3$ is commensurate and
without discommensurations, whereas that of $x=0.275$ is
incommensurate and does have discommensurations.

This leaves us one remaining feature of the spin excitation
spectrum to consider, namely the low energy diffuse scattering
distributed around the magnetic ordering wavevectors, shown in
Fig.\ \ref{fig:ridge}. As mentioned earlier, this scattering has a
different temperature dependence to the spin wave scattering, and
from this we can conclude that it arises from a different magnetic
component to the ordered spins. Low energy diffuse scattering was
also observed in the spin excitation spectrum of
La$_{5/3}$Sr$_{1/3}$NiO$_{4}$.\cite{boothroyd-PRL-2003} In that
case the diffuse scattering was almost one-dimensional and, like
the present case, there was no static component.

We attributed the quasi-1D diffuse scattering from
La$_{5/3}$Sr$_{1/3}$NiO$_{4}$ to dynamic AFM correlations among
the $S=\frac{1}{2}$ spins attached to lines of Ni$^{3+}$ holes in
the charge stripes, and the origin of the diffuse scattering from
La$_{3/2}$Sr$_{1/2}$NiO$_{4}$ is probably the same. Because the
Ni$^{3+}$ holes are arranged on a near-checkerboard pattern in
La$_{3/2}$Sr$_{1/2}$NiO$_{4}$ the spin correlations among the
Ni$^{3+}$ sites are expected to be quasi-2D, consistent with the
observed diffuse scattering. The slight elongation of the
scattering distribution is consistent with the introduction of a
stripe-like texture into the checkerboard pattern by
discommensurations. The width of the peak shown in Fig.\
\ref{fig:ridge}(c) is roughly twice that of an equivalent cut
through the diffuse scattering from La$_{5/3}$Sr$_{1/3}$NiO$_{4}$.
This indicates that the correlations between the Ni$^{3+}$ spins
are weaker in La$_{3/2}$Sr$_{1/2}$NiO$_{4}$. The interpretation of
the diffuse scattering in La$_{3/2}$Sr$_{1/2}$NiO$_{4}$ presented
here implies that there exists an AFM coupling between the
Ni$^{3+}$ spins that generates short-range fluctuations towards a
checkerboard ordering. The absence of static magnetic order on the
Ni$^{3+}$ sites could have implications for recent predictions of
orbital ordering in
La$_{3/2}$Sr$_{1/2}$NiO$_{4}$.\cite{Hotta-PRL-2004}

Finally, it is worth comparing our results with those obtained on
other checkerboard charge-ordered compounds. To our knowledge, the
only other such compound whose spin excitation spectrum has been
measured in detail is
La$_{3/2}$Sr$_{1/2}$CoO$_{4}$.\cite{helme-PhysicaB-2004}  The
spin-charge order is much closer to a perfect checkerboard pattern
in the half-doped cobaltate than in the half-doped nickelate. The
incommensurability observed at low temperatures for
La$_{3/2}$Sr$_{1/2}$CoO$_{4}$ is $\epsilon =
0.49$,\cite{Zaliznyak-PRL-2000} compared with $\epsilon = 0.44$
for La$_{3/2}$Sr$_{1/2}$NiO$_{4}$. Recall that for a perfect
checkerboard order $\epsilon = 0.5$. The measured spin excitation
spectrum of La$_{3/2}$Sr$_{1/2}$CoO$_{4}$ exhibits a simple spin
wave dispersion extending up 16 meV, and a second mode at energies
around 30\,meV which is relatively flat.\cite{helme-PhysicaB-2004}
The spin wave dispersion does not exhibit any of the unusual
features found in the case of La$_{3/2}$Sr$_{1/2}$NiO$_{4}$. This
can be understood if, as argued above, the deviations from a
simple spin wave picture are due to discommensurations: the very
small incommensurability of the cobaltate implies very few
discommensurations are present. The absence of low energy diffuse
scattering from the cobaltate also makes sense since the Co$^{3+}$
ions are believed to have a non-magnetic singlet ground
state.\cite{Zaliznyak-PRL-2000}\\

\section{\label{sec:conc}Conclusions}

The spin excitation spectrum of La$_{3/2}$Sr$_{1/2}$NiO$_{4}$ has
been found to contain a number of interesting features. We have
argued that the low energy diffuse scattering, which resembles a
similar signal previously observed in measurements on
La$_{5/3}$Sr$_{1/3}$NiO$_{4}$, originates from antiferromagnetic
correlations among spins attached to the Ni$^{3+}$ ions. We have
also argued that other strange features in the excitation spectrum
of La$_{3/2}$Sr$_{1/2}$NiO$_{4}$, such as the probable magnetic
modes of scattering dispersing from $(0.5,0.5)$ and $(1,0)$ type
positions and the large intrinsic widths of the spin excitations,
can be understood (at least at the level of hand waving) in terms
of a discommensuration model. The main obstacle in the way of a
more quantitative account of the spin excitation spectrum of
La$_{3/2}$Sr$_{1/2}$NiO$_{4}$ is the lack of a detailed model for
the ground state order.

We would like to thank L. M. Helme for help in calculating the
linear spin-wave dispersion, and Jos\'{e} Lorenzana for
stimulating discussions. This work was supported in part by the
Engineering and Physical Sciences Research Council of Great
Britain.


\begin{references}
\bibitem{notation}
In this work we describe the structural properties of
La$_{2-x}$Sr$_x$NiO$_{4+\delta}$ with reference to a tetragonal
unit cell with cell parameters $a \approx 3.8\,{\rm \AA}$ and $c
\approx 12.7\,{\rm \AA}$.
\end{references}
\end{document}